\documentclass[a4paper]{article}

\usepackage[margin=1in]{geometry}

\usepackage[T1]{fontenc}
\newcommand{\changefont}[3]{
\fontfamily{#1} \fontseries{#2} \fontshape{#3} \selectfont}

\changefont{ptm}{m}{n}

\usepackage{setspace} 
 \usepackage{graphicx}    

\newcommand \be{\begin{equation}}
\newcommand \ee{\end{equation}}
\newcommand \ba{\begin{eqnarray}}
\newcommand \ea{\end{eqnarray}}

\def\bit{\begin{itemize}}
\def\eit{\end{itemize}}

\usepackage{amsfonts}
\usepackage{amssymb}
\usepackage{graphics}
\usepackage{mathrsfs}
\usepackage{color}

\newtheorem{theorem}{Theorem}[section]

\newtheorem{lemma}{Lemma}[section]

\newtheorem{definition}{Definition}[section]

\long\def\symbolfootnote[#1]#2{\begingroup%
\def\thefootnote{\fnsymbol{footnote}}\footnote[#1]{#2}\endgroup} 

\begin{document}

%

\begin{center}
\Large \textbf{Li-Yorke chaos in hybrid systems on a time scale}
\end{center}

\vspace{-0.3cm}
\begin{center}
\normalsize \textbf{Marat Akhmet$^{a,} \symbolfootnote[1]{Corresponding Author Tel.: +90 312 210 5355,  Fax: +90 312 210 2972, E-mail: marat@metu.edu.tr}$, Mehmet Onur Fen$^b$} \\
\vspace{0.2cm}
\textit{\textbf{\footnotesize$^a$Department of Mathematics, Middle East Technical University, 06800, Ankara, Turkey}} \\
\textit{\textbf{\footnotesize$^b$Neuroscience Institute, Georgia State University, Atlanta, Georgia 30303, USA}}
\vspace{0.1cm}
\end{center}

\vspace{0.3cm}

\begin{center}
\textbf{Abstract}
\end{center}

\noindent\ignorespaces

By using the reduction technique to impulsive differential equations \cite{Akhmet06}, we rigorously prove the presence of chaos in dynamic equations on time scales (DETS). The results of the present study are based on the Li-Yorke definition of chaos. This is the first time in the literature that chaos is obtained for DETS. An illustrative example is presented by means of a Duffing equation on a time scale.

\vspace{0.2cm}
 
\noindent\ignorespaces \textbf{Keywords:} Li-Yorke chaos; Dynamic equations on time scales; Proximality; Frequent separation; Duffing equation; Hybrid systems

\section{Introduction}\label{timescale_intro}

The concept of chaos has been one of the attractive topics among scientists since the studies of Poincar\'{e} \cite{Andersson94}, Cartwright and Littlewood \cite{Cartwright1}, Levinson \cite{Levinson}, Lorenz \cite{Lorenz63} and Ueda \cite{Ueda78}. Another subject that is also popular is the theory of time scales, which is first presented by Hilger \cite{Hilger88}. Both concepts have many applications in various disciplines such as mechanics, electronics, neural networks, population models and economics. See, for instance, \cite{Barrio14,Bohner01,Feckan11,Grebogi97,AlbertLuo,Owens13,Thamilmaran04,Tisdell08,Zhang10} and the references therein.

Dynamic equations on time scales (DETS) have been extensively investigated in the literature \cite{Bohner01,Lakshmikantham96}. However, to the best of our knowledge, the presence of chaos has never been achieved in DETS. Motivated by the deficiency of mathematical methods for the investigation of chaos in such equations, we suggest the results of the present study.

The first mathematical definition of chaos was introduced by Li and Yorke \cite{Li75} for discrete dynamical systems in a compact interval of the real line. The presence of an uncountable scrambled set is one of the main features of the Li-Yorke chaos. The original definition of Li and Yorke was extended to dimensions greater than one by Marotto \cite{Marotto78}. According to Marotto \cite{Marotto78}, a multidimensional continuously differentiable map possesses generalized Li-Yorke chaos if it has a snap-back repeller. The existence of Li-Yorke chaos in a spatiotemporal chaotic system was proved in \cite{PLi07} by means of Marotto's Theorem, and generalizations of Li-Yorke chaos to mappings in Banach spaces and complete metric spaces were provided in \cite{Kloeden06,Shi04,Shi05}. It was shown by Kuchta \cite{Kuchta} that if a map on a compact interval has a two point scrambled set, then it possesses an uncountable scrambled set. Blanchard \cite{Blanchard02} proved that the presence of positive topological entropy implies chaos in the sense of Li-Yorke. Moreover, Li-Yorke chaos on several spaces in connection with the cardinality of its scrambled sets was studied within the scope of the paper \cite{Guirao05}. Besides, Li-Yorke sensitivity, which links the Li-Yorke chaos with the notion of sensitivity, was studied in the paper \cite{Akin03}. The studies \cite{Akh5,Akh2,Akh4,Akh7,Akh9,Akh8,Akh14} were concerned with the extension of chaos in continuous-time systems that possess asymptotically stable and hyperbolic equilibria as well as orbitally stable limit cycles. It was found in these papers that the solutions admit the same type of chaos as the perturbations. The paper \cite{Akh4} deals with the general technique of dynamical synthesis, which was developed in \cite{Brown93}--\cite{Brown01}. In the present study, we develop the concept of Li-Yorke chaos for DETS and prove its existence rigorously. Our results are appropriate to obtain chaotic DETS with arbitrary high dimensions.
 
Throughout the paper, we will denote by $\mathbb R,$ $\mathbb Z$ and $\mathbb N$ the sets of real numbers, integers and natural numbers, respectively. In this study, we consider the following equation,
\begin{eqnarray} \label{main_eqn1}
y^{\Delta} (t)=Ay(t) + f(t,y(t)) + g(t,\zeta), ~t\in\mathbb{T}_0 ,
\end{eqnarray}
where  $A$ is a constant $n\times n$ real valued matrix, the function $f:\mathbb T_0 \times \mathbb R^n \to \mathbb R^n$ is rd-continuous and the function $g(t,\zeta)$ is defined through the equation $g(t,\zeta)=\zeta_k$ for $t\in [\theta_{2k-1}, \theta_{2k}],$ $k\in\mathbb Z,$ such that $\zeta=\left\{\zeta_k\right\}$ is a sequence generated by the map 
\begin{eqnarray}\label{timescale_mapF}
\zeta_{k+1}=F(\zeta_k),
\end{eqnarray}
where $\zeta_0 \in \Lambda,$  $F:\Lambda \to \Lambda$ is a continuous function and $\Lambda$ is a compact subset of $\mathbb R^n.$ 
In equation (\ref{main_eqn1}) the time scale $\mathbb T_0$ is defined as $\mathbb T_0 = \bigcup_{k=-\infty}^{\infty} [\theta_{2k-1}, \theta_{2k}]$ in which $\left\{\theta_k\right\}$ is a strictly increasing sequence of real numbers such that $\left|\theta_k\right| \to \infty$ as $\left|k\right|\to \infty$ and $\sum_{-\infty} (\theta_{2k}-\theta_{2k-1}) =\infty,$ $\sum^{\infty} (\theta_{2k}-\theta_{2k-1})=\infty.$ 

In the present paper, we investigate the existence of chaos in the dynamics of equation (\ref{main_eqn1}). The system under discussion is a hybrid one, since it combines the continuous dynamics on the time scale with the discrete equation used in the right hand side of the system. We theoretically prove that chaos exists in (\ref{main_eqn1}) provided that the map (\ref{timescale_mapF}) is chaotic. For that purpose, we make use of the reduction technique to impulsive differential equations, which was presented by Akhmet and Turan \cite{Akhmet06}. As far as we know, there is no paper on chaos in dynamics on time scales. The reason is that the dynamics is essentially non-autonomous and it is difficult to verify the ingredients of chaos for unspecified time scales. That is why we utilize the time scale introduced in the papers \cite{Akhmet06,Akhmet09} and the method of reduction of the dynamics to impulsive differential equations \cite{Akhmet06}.

The rest of the paper is organized as follows. In Section \ref{time_scales_sec2}, some preliminary results as well as basic concepts about DETS are mentioned. Section \ref{time_scale_bounded} is devoted to the bounded solutions of (\ref{main_eqn1}). In Section \ref{section3_timescalepaper}, we give the description of the chaos of equation (\ref{main_eqn1}) and prove its presence rigorously. An example concerning Duffing equations on a time scale is presented in Section \ref{example_time_scales} to support the theoretical results. Finally, some concluding remarks are given in Section \ref{time_scales_conclusion}.

\section{Preliminaries} \label{time_scales_sec2}

The basic concepts that are needed in the present paper about differential equations on time scales are as follows \cite{Bohner01,Lakshmikantham96,Lakshmikantham02,Lakshmikantham06}. A time scale is a nonempty closed subset of $\mathbb{R}.$ On a time scale $\mathbb{T}$, the forward and backward jump operators are defined as $\sigma(t)=\inf\left\{s\in\mathbb{T}: s>t \right\}$ and $\rho(t)=\sup\left\{s\in\mathbb{T}: s<t\right\},$ respectively. We say that a point $t\in\mathbb{T}$ is right-scattered if $\sigma(t)>t$ and right-dense if $\sigma(t)=t$. In a similar way, if $\rho(t)<t,$ then $t\in\mathbb{T}$ is called left-scattered, and otherwise it is called left-dense. Besides, a function $h:\mathbb{T} \times \mathbb{R}^n \to \mathbb{R}^n$ is called rd-continuous if it is continuous at each $(t,u) \in \mathbb{T} \times \mathbb{R}^n$ with right-dense $t,$ and the limits $\displaystyle \lim_{(r,\nu) \to (t^-,u)} h(r,\nu) =h(t-,u)$ and $\displaystyle \lim_{\nu \to u} h(t,\nu)=h(t,u)$ exist at each $(t,u)$ with left-dense $t.$ At a right-scattered point $t\in\mathbb{T},$ the $\Delta$-derivative of a continuous function $\vartheta$ is defined as $\vartheta^{\Delta}\left(t\right)=\displaystyle \frac{\vartheta\left(\sigma\left(t\right)\right)-\vartheta\left(t\right)}{\sigma\left(t\right)-t}.$ On the other hand, at a right-dense point $t,$ we have $\vartheta^{\Delta}\left(t\right)=\displaystyle \lim_{r \to t}\frac{\vartheta\left(t\right)-\vartheta\left(r\right)}{t-r}$ provided that the limit exists. 

It is worth noting that on the time scale $\mathbb T_0$ used in system (\ref{main_eqn1}) the points $\theta_{2k-1}$, $k\in\mathbb{Z},$ are left-scattered and right-dense, and the points $\theta_{2k}$, $k\in\mathbb{Z},$ are right-scattered and left-dense. Moreover, $\sigma(\theta_{2k})=\theta_{2k+1}$, $\rho(\theta_{2k+1})=\theta_{2k}$, $k\in\mathbb{Z},$ and $\sigma(t)=\rho(t)=t$ for any $t\in\mathbb{T}_{0}$ except at the points $\theta_k,$ $k\in\mathbb Z.$

Suppose that the time scale $\mathbb T_0$ used in the description of equation (\ref{main_eqn1}) satisfies the $\omega-$property. That is, there exists a number $\omega >0$ such that $t+\omega \in \mathbb T_0$ whenever $t \in \mathbb T_0.$ In this case, there exists a natural number $p$ such that $\delta_{k+p}=\delta_k$ for all $k\in \mathbb Z,$ where $\delta_k=\theta_{2k+1}-\theta_{2k}$ \cite{Akhmet06}. Suppose that $p$ is the minimal among those numbers.

We assume without loss of generality that $\theta_{-1}<0<\theta_0.$ Define on the set $\mathbb T'_0=\mathbb T_0 \setminus \bigcup_{k =-\infty}^{\infty} \left\{\theta_{2k-1} \right\}$ the $\psi-$substitution \cite{Akhmet06} as
\begin{eqnarray} \label{func_psi}
\displaystyle \psi(t)=\left\{\begin{array}{ll} \displaystyle t-\sum_{0<\theta_{2k}<t} \delta_k,    &   t  \ge 0, \\
                                                 \displaystyle t+\sum_{t\le\theta_{2k}<0} \delta_k,    &   t < 0, 
\end{array} \right..
\end{eqnarray}
The function $\psi(t)$ is  one-to-one, $\psi(0)=0,$ $\psi(\mathbb T'_0)=\mathbb R$ and $\displaystyle \lim_{t\to\infty, ~t\in \mathbb T_0'}\psi(t) = \infty.$ According to the results of the paper \cite{Akhmet06}, $d\psi(t)/dt=1,$ $t\in \mathbb T'_0,$ and $d\psi^{-1}(s)/ds=1$ provided that $s\neq s_k,$ $k\in\mathbb Z,$ where 
\begin{eqnarray} \label{func_psi_inv}
\displaystyle \psi^{-1}(s)=\left\{\begin{array}{ll} \displaystyle s+\sum_{0<s_{k}<s} \delta_k,    &   s  \ge 0, \\
                                                 \displaystyle s-\sum_{s\le s_{k}<0} \delta_k,    &   s < 0, 
\end{array} \right.
\end{eqnarray}
and the sequence $\left\{s_k\right\},$ $k\in \mathbb Z,$ is defined through the equation $s_k=\psi(\theta_{2k}).$ The function $\psi^{-1}$ is piecewise continuous with discontinuities of the first kind at the points $s_k,$ $k\in \mathbb Z,$ such that $\psi^{-1}(s_k+)-\psi^{-1}(s_k)=\delta_k,$ where $\psi^{-1}(s_k+)=\displaystyle \lim_{s\to s_k^+} \psi^{-1}(s),$ the sequence $\left\{s_k\right\}$ is $(\psi(\omega),p)-$periodic, i.e., $s_{k+p}=s_k+\psi(\omega)$ for all $k\in\mathbb Z,$ and $\psi(t+\omega)=\psi(t)+\psi(\omega),$ $t\in \mathbb T'_0.$ Moreover, if a function $h(t)$ is $\omega-$periodic on $\mathbb T_0,$ then $h(\psi^{-1}(s))$ is $\psi(\omega)-$periodic, and vice versa.

Let us denote by ${\cal C}_{rd}(\mathbb T_0)$ the set of all functions which are rd-continuous on $\mathbb T_0,$ and let ${\cal C}^1_{rd}(\mathbb T_0)\subset {\cal C}_{rd}(\mathbb T_0)$ be the set of all continuously differentiable functions on $\mathbb T_0,$ assuming that the functions have a one sided derivative at $\theta_k,$ $k\in\mathbb Z.$ On the other hand, we say that a function defined on $\mathbb R$ is an element of the set ${\cal PC}_0$ if it is left-continuous on $\mathbb R$ and continuous on $\mathbb R \setminus \bigcup_{k=-\infty}^{\infty}\left\{s_k\right\},$ and it has discontinuities of the first kind at the points $s_k,$ $k\in\mathbb Z.$ Moreover, a function $h:\mathbb R \to \mathbb R^n$ belongs to the set ${\cal PC}^1_0$ if both $h$ and $h'$ are elements of ${\cal PC}_0,$ where $h'(s_k)= \displaystyle \lim_{s\to s_k^-} \frac{h(s)-h(s_k)}{s-s_k},$ $k\in \mathbb Z.$ It was shown by Akhmet and Turan \cite{Akhmet06} that a function $\vartheta(t)$ belongs to ${\cal C}_{rd}(\mathbb T_0)$ $\left({\cal C}^1_{rd}(\mathbb T_0)\right)$ if and only if $\vartheta(\psi^{-1}(s))$ belongs to ${\cal PC}_0$ $\left({\cal PC}^1_0\right).$  

In accordance with the equation $\displaystyle y^{\Delta} (\theta_{2k}) = \frac{y(\theta_{2k+1})-y(\theta_{2k})}{\theta_{2k+1}-\theta_{2k}},$ $k\in \mathbb Z,$ system (\ref{main_eqn1}) can be written as
\begin{eqnarray} \label{main_eqn2}
\begin{array}{l}
y' (t)=Ay(t) + f(t,y(t)) + g(t,\zeta), ~t\in\mathbb{T}_0, \\
y(\theta_{2k+1}) =  \delta_k  A  y(\theta_{2k}) + f(\theta_{2k},y(\theta_{2k})) \delta_k + \zeta_k\delta_k + y(\theta_{2k}).
\end{array}
\end{eqnarray}
Applying the transformation $s=\psi(t)$ to (\ref{main_eqn2}) we obtain the following impulsive system,
\begin{eqnarray} \label{main_eqn3}
\begin{array}{l}
x'(s)=Ax(s) + f(\psi^{-1}(s),x(s)) + g(\psi^{-1}(s),\zeta), ~s\neq s_k, \\
\displaystyle \Delta x |_{s=s_k} =  \delta_k A  x(s_k) + f(\psi^{-1}(s_k),x(s_k)) \delta_k +\zeta_k\delta_k,
\end{array} 
\end{eqnarray}
where $x(s)=y(\psi^{-1}(s)),$ $\displaystyle \Delta x |_{s=s_k}=x(s_k+)-x(s_k),$ $k\in \mathbb Z,$ and $\displaystyle x(s_k+)=\lim_{s\to s_k^+} x(s).$

In what follows, we will make use of the usual Euclidean norm for vectors and the norm induced by the Euclidean norm for square matrices \cite{Horn92}.

The following conditions are required throughout the paper. 
\begin{enumerate}
\item[\textbf{(C1)}] $\det(I + \delta_kA) \neq 0$ for all $k\in \mathbb Z,$ where $I$ is the $n \times n$ identity matrix;
\item[\textbf{(C2)}] All eigenvalues of the matrix $e^{\psi(\omega)A} \displaystyle \Pi_{j=0}^{p-1} (I  + \delta_j A)$ lie inside the unit circle; 
\item[\textbf{(C3)}] There exist positive numbers $M_f$ and $M_F$ such that $\displaystyle \sup_{t\in \mathbb T_0, ~y\in \mathbb R^n} \left\|f(t,y)\right\| \le M_f$ and $\displaystyle \sup_{\eta \in \Lambda} \left\| F(\eta)  \right\| \le M_F;$
\item[\textbf{(C4)}] There exists a positive number $L_f$ such that $\left\|f(t,y_1)-f(t,y_2)\right\| \le L_f \left\|y_1-y_2\right\|$ for all $t\in \mathbb T_0$ and $y_1,y_2\in \mathbb R^n.$ 
\end{enumerate}

Let us denote by $X(s,r)$ the transition matrix of the linear homogeneous system
\begin{eqnarray} 
\begin{array}{l}
x'(s)=Ax(s), ~s \neq s_k,\\
\Delta x|_{s=s_k} =  \delta_k A x(s_k).
\end{array}
\end{eqnarray} 
Under the conditions $(C1)$ and $(C2)$ there exist positive numbers $N$ and $\lambda$ such that $\left\|X(s,r)\right\| \le Ne^{-\lambda (s-r)}$ for $s\ge r$ \cite{Akh1,Samolienko95}.

The following conditions are also needed. 
\begin{enumerate}
\item[\textbf{(C5)}] $\displaystyle N L_f\left( \frac{1}{\lambda} + \frac{p \bar{\delta}}{1-e^{-\lambda \psi(\omega)}}  \right)<1,$ where $\displaystyle \bar{\delta}=\max_{0\le k \le p-1}\delta_k;$
\item[\textbf{(C6)}] $\displaystyle -\lambda+NL_f+ \frac{p}{\psi(\omega)} \ln \left(1+NL_f\bar{\delta}\right)<0;$ 
\item[\textbf{(C7)}] $f(t+\omega,y)=f(t,y)$ for all $(t,y)\in \mathbb T_0 \times \mathbb R^n.$
\end{enumerate}

The next section is devoted to the bounded solutions of system (\ref{main_eqn1}).

\section{Bounded solutions} \label{time_scale_bounded}

Under the conditions $(C1)-(C5),$ one can verify by using the results of \cite{Akh1,Samolienko95} that for a fixed sequence $\zeta=\left\{\zeta_k\right\},$ $k\in \mathbb Z,$ there exists a unique bounded on $\mathbb R$ solution $\phi_{\zeta}(s)$ of (\ref{main_eqn3}), which satisfies the relation
\begin{eqnarray}
\begin{array}{l}
\phi_{\zeta}(s)= \displaystyle \int_{-\infty}^s X(s,r) \left[  f\left(\psi^{-1}(r), \phi_{\zeta}(r)\right) + g\big(\psi^{-1}(r),\zeta \big)  \right] dr \\
+ \displaystyle \sum_{-\infty < s_k < s} X(s,s_k+) \left[ f\big(\psi^{-1}(s_k),\phi_{\zeta}(s_k)\big) + \zeta_k \right] \delta_k.
\end{array}
\end{eqnarray}
Moreover, $\displaystyle \sup_{s\in\mathbb R} \left\|\phi_{\zeta}(s)\right\| \le \displaystyle K_0,$ where $K_0=N(M_f+M_F) \Big(\displaystyle\frac{1}{\lambda}+\displaystyle\frac{p\bar{\delta}}{1-e^{-\lambda \psi(\omega)}}\Big).$ Therefore, for a fixed sequence $\zeta=\left\{\zeta_k\right\},$ the function $\varphi_{\zeta}(t)=\phi_{\zeta}(\psi(t))$ satisfying $\varphi_{\zeta}(\theta_{2k+1})=\phi_{\zeta}(s_k+),$ $k\in\mathbb Z,$ is the unique solution of (\ref{main_eqn2}), and hence of (\ref{main_eqn1}), which is bounded on $\mathbb T_0$ such that $\displaystyle \sup_{t\in\mathbb T_0} \left\|\varphi_{\zeta}(t)\right\| \le K_0.$

We say that the bounded solution $\varphi_{\zeta}(t)$ attracts a solution  $y(t)$ of (\ref{main_eqn1}) if $\left\|y(t)-\varphi_{\zeta}(t)\right\|\to 0$ as $t\to \infty,$ $t\in\mathbb T_0.$ The attractiveness feature of the bounded solutions of (\ref{main_eqn1}) is mentioned in the next assertion.

\begin{lemma}\label{attractiveness_timescalepaper}
If the conditions $(C1)-(C6)$ are valid, then for a fixed sequence $\zeta,$ the bounded solution $\varphi_{\zeta}(t)$ attracts all other solutions of (\ref{main_eqn1}).
\end{lemma}

\noindent \textbf{Proof.} Consider an arbitrary solution  $y(t),$ $y(t^0)=y_0,$ of (\ref{main_eqn1}) for some $t^0 \in \mathbb T_0$ and $y_0\in \mathbb R^n.$ Assume without loss of generality that $t^0\neq \theta_{2k-1}$ for any $k\in\mathbb Z.$ Let $s^0=\psi(t^0)$ and $x(s)=y(\psi^{-1}(s)).$ The relation
\begin{eqnarray*}
&& x(s) - \phi_{\zeta}(s) = X(s,s^0) (y_0 -\phi_{\zeta}(s^0)) + \displaystyle \int_{s^0}^s \left[ f\big(\psi^{-1}(r),x(r)\big) - f\big(\psi^{-1}(r),\phi_{\zeta}(r)\big)  \right] dr \\
&& + \displaystyle \sum_{s^0 \le s_k < s} X(s,s_k+) \left[  f\big(\psi^{-1}(s_k),x(s_k)\big)  -  f\big(\psi^{-1}(s_k),\phi_{\zeta}(s_k)\big)  \right] \delta_k
\end{eqnarray*}
implies for $s\ge s^0$ that
\begin{eqnarray*}
&& \left\| x(s) - \phi_{\zeta}(s)\right\| \le Ne^{-\lambda (s-s^0)} \left\|y_0 -\phi_{\zeta}(s^0)\right\| +  \displaystyle \int_{s^0}^s NL_f e^{-\lambda (s-r)} \left\| x(r) - \phi_{\zeta}(r)\right\| dr \\
&& + \displaystyle \sum_{s^0 \le s_k < s} NL_f \bar{\delta}  e^{-\lambda (s-s_k)} \left\| x(s_k) - \phi_{\zeta}(s_k)\right\|. 
\end{eqnarray*}
Applying the Gronwall-Bellman Lemma for piecewise continuous functions \cite{Akh1} to the last inequality, one can obtain   that 
$$
\left\| x(s) - \phi_{\zeta}(s)\right\| \le N (1+ NL_f \bar{\delta})^p \left\| y_0 - \phi_{\zeta} (s^0)  \right\| e^{[-\lambda + NL_f + p\ln (1+NL_f\bar{\delta})/ \psi(\omega)](s-s^0)}, ~s\ge s^0. 
$$
Therefore, we have for $t\ge t^0,$ $t\in \mathbb T_0,$ that
$$
\left\| y(t) - \varphi_{\zeta}(t)\right\| \le N (1+ NL_f \bar{\delta})^p \left\| y_0 - \varphi_{\zeta} (t^0)  \right\| e^{[-\lambda + NL_f + p\ln (1+NL_f\bar{\delta})/ \psi(\omega)](\psi(t)-\psi(t^0))}. 
$$
Consequently, $\left\| y(t) - \varphi_{\zeta}(t)\right\| \to 0$ as $t \to \infty,$ $t\in \mathbb T_0.$ $\square$  

In the next section, we will deal with the presence of chaos in system (\ref{main_eqn1}).

\section{The chaotic dynamics} \label{section3_timescalepaper}

The map (\ref{timescale_mapF}) is called Li-Yorke chaotic on $\Lambda$ if \cite{Akin03,Aulbach01,Kolyada04,Li75,Li93}: 
(i) For every natural number $p_0,$ there exists a $p_0-$periodic point of $F$ in $\Lambda;$ 
(ii) There is an uncountable set ${\mathcal S} \subset \Lambda,$ the scrambled set, containing no periodic points, such that for every $\zeta^{1},\zeta^{2}\in {\mathcal S}$ with $\zeta^{1}\neq \zeta^{2},$ we have $\displaystyle \limsup_{k\to\infty}\big\|F^k(\zeta^1)-F^k(\zeta^2)\big\|>0$ and $\displaystyle \liminf_{k\to\infty}\big\|F^k(\zeta^1)-F^k(\zeta^2)\big\|=0;$ 
(iii) For every $\zeta^1\in {\mathcal S}$ and a periodic point $\zeta^2 \in \Lambda,$ we have $\displaystyle \limsup_{k\to\infty}\big\|F^k(\zeta^1)-F^k(\zeta^2)\big\|>0.$

Let us denote by $\Theta$ the set of all sequences $\zeta=\left\{\zeta_k\right\},$ $k\in \mathbb Z,$ obtained by equation (\ref{timescale_mapF}). A pair of sequences $\zeta=\left\{\zeta_k\right\},$ $\tilde{\zeta}=\left\{\tilde{\zeta}_k\right\}\in\Theta$ is proximal if $\displaystyle \liminf_{k\rightarrow\infty} \big\|\zeta_{k}-\tilde{\zeta}_{k} \big\|=0.$ Moreover, the pair is frequently separated if $\displaystyle \limsup_{k\rightarrow\infty} \big\|\zeta_{k}-\tilde{\zeta}_{k}\big\| >0.$

We say that a pair $\varphi_{\zeta}(t)$, $\varphi_{\tilde{\zeta}}(t)$ of bounded solutions of (\ref{main_eqn1}) is proximal if for an arbitrary small real number $\epsilon>0$ and arbitrary large natural number $E,$ there exists an integer $m$ such that $\big\| \varphi_{\zeta}(t)- \varphi_{\tilde{\zeta}}(t)\big\| <\epsilon$ for all $t\in [\theta_{2m-1}, \theta_{2(m+E)}] \cap \mathbb T_0.$  
On the other hand, the pair $\varphi_{\zeta}(t)$, $\varphi_{\tilde{\zeta}}(t)$ is frequently $(\epsilon_{0},\Delta)$-separated if there exist  numbers $\epsilon_{0}>0,$ $\Delta>0$ and infinitely many disjoint intervals $J_q \subset \mathbb T_0,$ $q\in\mathbb N,$ each with a length no less than $\Delta,$ such that $\big\| \varphi_{\zeta}(t) - \varphi_{\tilde{\zeta}}(t)\big\| >\epsilon_{0}$ for each $t$ from these intervals.
Furthermore, a pair  $\varphi_{\zeta}(t)$, $\varphi_{\tilde{\zeta}}(t)$ of solutions of (\ref{main_eqn1}) is called a Li-Yorke pair if it is proximal and frequently $(\epsilon_{0},\Delta)$-separated for some positive numbers $\epsilon_{0}$ and $\Delta.$ 

Let $\mathscr{A}$ be the collection of all bounded solutions $\varphi_{\zeta}(t)$ of (\ref{main_eqn1}) such that $\zeta \in \Theta.$ The description of Li-Yorke chaos for system (\ref{main_eqn1}) is as follows.

\begin{definition} \label{timescale_chaos_defn}
System (\ref{main_eqn1}) is called Li-Yorke chaotic if:
\begin{enumerate}
\item [(i)] There exists an $m\omega-$periodic solution of (\ref{main_eqn1}) for each $m\in \mathbb N;$  
\item  [(ii)] There exists an uncountable set $\Sigma\subset \mathscr{A},$ the scrambled set, which does not contain any periodic solution,  such that any pair of different solutions of (\ref{main_eqn1}) inside $\Sigma$ is a Li-Yorke pair; 
\item [(iii)] For any $\varphi_{\zeta}(t)\in \Sigma$ and any periodic solution $\varphi_{\hat{\zeta}}(t)\in \mathscr{A},$ the pair $\varphi_{\zeta}(t)$, $\varphi_{\hat{\zeta}}(t)$ is frequently $(\epsilon_{0},\Delta)$-separated for some positive numbers $\epsilon_{0}$ and $\Delta.$
\end{enumerate}
\end{definition}

One can verify that the sequence $\left\{\kappa_k\right\}$ defined through the equation $\kappa_k=\theta_{2k}-\theta_{2k-1},$ $k\in \mathbb Z,$ is $p$-periodic. In what follows, we will denote $\displaystyle \underline{\kappa} = \min_{0\le k \le p-1} \kappa_k$ and $\displaystyle \overline{\kappa} = \max_{0\le k \le p-1} \kappa_k.$ Moreover, let $i((a_0,b_0))$ be the number of the terms of the sequence $\left\{s_k\right\}$ that belong to the interval $(a_0,b_0),$ where $a_0, b_0 \in \mathbb R$ with $a_0< b_0.$ One can verify that $i((a_0,b_0)) \le p + \displaystyle \frac{p}{\psi(\omega)} (b_0-a_0).$

The next assertion is about the proximality feature of bounded solutions of equation (\ref{main_eqn1}).

\begin{lemma} \label{timescale_prox}
Suppose that the conditions $(C1)-(C6)$ are fulfilled. If a pair of sequences $\zeta,\tilde{\zeta}\in\Theta$ is proximal, then the same is true for the pair $ \varphi_{\zeta}(t), \varphi_{\tilde{\zeta}}(t)\in\mathscr{A}.$
\end{lemma}

\noindent  \textbf{Proof.}
Set $\displaystyle R_1=2N(M_f+M_F) \Big( \frac{1}{\lambda} + \frac{p\bar{\delta}}{1-e^{-\lambda \psi(\omega)}} \Big)$ and $\alpha=\lambda - NL_f-\displaystyle \frac{p}{\psi(\omega)} \ln(1+NL_f \bar{\delta}).$ Suppose that $\gamma$ is a real number which satisfies the inequality 
$$
\gamma \ge \displaystyle  1+ N\Big(  \frac{1}{\lambda} + \frac{\bar{\delta} p}{1-e^{-\lambda \psi(\omega)}}  \Big) \Big( 1+  \frac{NL_f(1+NL_f\bar{\delta})^p}{\alpha} + \frac{NL_f\bar{\delta}p(1+NL_f\bar{\delta})^p }{1-e^{-\alpha \psi(\omega)}} \Big).
$$
Fix an arbitrary small number $\epsilon>0$ and an arbitrary large natural number $E$ such that $$\displaystyle E\ge \frac{1}{\alpha \underline{\kappa}} \ln\Big( \frac{\gamma R_1(1+NL_f \bar{\delta})^p}{\epsilon} \Big).$$ Since the pair $\zeta,$ $\tilde{\zeta}$ is proximal, there exists an integer $k_0$ such that $\left\| g\big(t,\zeta\big)-g\big(t,\tilde{\zeta}\big) \right\|<\epsilon / \gamma$ for $t\in [\theta_{2k_0-1}, \theta_{2(k_0+2E)}]\cap \mathbb T_0.$ In this case, $\left\| g\big(\psi^{-1}(s),\zeta\big)-g\big(\psi^{-1}(s),\tilde{\zeta}\big)  \right\|<\epsilon / \gamma$ for $s\in (s_{k_0-1}, s_{k_0+2E}].$

The bounded solutions $\phi_{\zeta}(s)=\varphi_{\zeta}(\psi^{-1}(s))$ and $\phi_{\tilde{\zeta}}(s)=\varphi_{\tilde{\zeta}}(\psi^{-1}(s))$ of (\ref{main_eqn3}) satisfy the relation
\begin{eqnarray*}
&& \phi_{\zeta}(s) - \phi_{\tilde{\zeta}}(s) = \displaystyle \int_{-\infty}^{s} X(s,r) \Big[ f \big(\psi^{-1}(r), \phi_{\zeta}(r) \big)-f \big(\psi^{-1}(r), \phi_{\tilde{\zeta}}(r) \big) \\
&&  + g \big(\psi^{-1}(r),\zeta \big)   - g \big(\psi^{-1}(r),\tilde{\zeta} \big)  \Big] dr \\
&& + \displaystyle \sum_{-\infty < s_k < s} X(s,s_k+) \Big[ f \big(\psi^{-1}(s_k),\phi_{\zeta}(s_k) \big) - f \big(\psi^{-1}(s_k),\phi_{\tilde{\zeta}}(s_k) \big)+ \zeta_k  - \tilde{\zeta}_k \Big] \delta_k.
\end{eqnarray*}
Thus, for $s\in (s_{k_0-1}, s_{k_0+2E}],$ we have that
\begin{eqnarray} \label{time_scale_prox_proof_1}
\begin{array}{l}
 \Big\| \phi_{\zeta}(s) - \phi_{\tilde{\zeta}}(s)  \Big\| \le R_1  e^{-\lambda(s- s_{k_0-1})} + \displaystyle \frac{N\epsilon}{\gamma\lambda} \left( 1- e^{-\lambda(s-s_{k_0-1})} \right)\\
 + \displaystyle \frac{N\bar{\delta}p\epsilon}{\gamma (1-e^{-\lambda \psi(\omega)})} \left(1 -e^{-\lambda(s-s_{k_0-1}+\psi(\omega))}  \right) \\
 + \displaystyle \int^{s}_{s_{k_0-1}} NL_f e^{-\lambda (s-r)} \Big\| \phi_{\zeta}(r) - \phi_{\tilde{\zeta}}(r) \Big\| dr  \\
 + \displaystyle \sum_{s_{k_0-1} < s_k < s } N L_f \bar{\delta} e^{-\lambda (s-s_k)} \Big\| \phi_{\zeta}(s_k) - \phi_{\tilde{\zeta}}(s_k)  \Big\|. 
\end{array}
\end{eqnarray} 
 
Let us define the functions $u(s)=e^{\lambda s} \Big\| \phi_{\zeta}(s) - \phi_{\tilde{\zeta}}(s)  \Big\|$ and $v(s)=\beta_1+\beta_2  e^{\lambda s},$ where 
$$
\beta_1= R_1 e^{\lambda s_{k_0-1}} - \frac{N\epsilon}{\gamma \lambda} e^{\lambda s_{k_0-1}} - \frac{N \bar{\delta} p \epsilon}{\gamma (1-e^{-\lambda \psi(\omega)})} e^{\lambda (s_{k_0-1}-\psi(\omega))}
$$
and
$$
\beta_2= \frac{N\epsilon}{\gamma}  \left(  \frac{1}{\lambda} + \frac{ \bar{\delta} p }{ 1-e^{-\lambda \psi(\omega)} }  \right).
$$
One can confirm by means of (\ref{time_scale_prox_proof_1}) that
\begin{eqnarray*}
u(s) \le v(s) + \displaystyle \int^s_{s_{k_0-1}} NL_f u(r) dr + \sum_{s_{k_0-1} < s_k < s } NL_f \bar{\delta} u(s_k).
\end{eqnarray*}
It can be shown by applying the analogue of the Gronwall's Lemma for piecewise continuous functions that
\begin{eqnarray*}
  u(s) \le v(s) + \displaystyle \int_{s_{k_0-1}}^s N L_f  (1+ N L_f \bar{\delta})^{i((r,s))} e^{NL_f (s-r)} v(r) dr \\
  + \displaystyle \sum_{s_{k_0-1} < s_k < s} N L_f \bar{\delta} (1+ N L_f \bar{\delta})^{i((s_k,s))} e^{NL_f (s-s_k)} v(s_k).
\end{eqnarray*}  
Accordingly, the inequality  
\begin{eqnarray*} 
&& u(s) \le \beta_1 (1+NL_f\bar{\delta})^p e^{(\lambda-\alpha) (s-s_{k_0-1})} + \beta_2 e^{\lambda s} \\
&& + \frac{NL_f \beta_2  (1+NL_f\bar{\delta})^p}{\alpha} e^{\lambda s} \left(1- e^{-\alpha (s-s_{k_0-1}) } \right) \\
&& + \frac{ NL_f\bar{\delta} p \beta_2 (1+NL_f\bar{\delta})^p }{1-e^{-\alpha \psi(\omega)}}  e^{\lambda s} \left(1- e^{-\alpha (s-s_{k_0-1}+\psi(\omega)) } \right).
\end{eqnarray*} 
is valid. Therefore,
\begin{eqnarray*} 
&& \Big\| \phi_{\zeta}(s) - \phi_{\tilde{\zeta}}(s)  \Big\| < R_1 (1+NL_f \bar{\delta})^p e^{-\alpha (s-s_{k_0-1})} \\
&& + \displaystyle \frac{N\epsilon}{\gamma} \left(  \frac{1}{\lambda} + \frac{\bar{\delta} p }{1-e^{-\lambda \psi(\omega)}}  \right) \left( 1+  \frac{NL_f(1+NL_f\bar{\delta})^p}{\alpha} + \frac{NL_f\bar{\delta}p(1+NL_f\bar{\delta})^p}{1-e^{-\alpha \psi(\omega)}} \right)
\end{eqnarray*}  
for  $s \in (s_{k_0-1},s_{k_0+2E}].$
 
Suppose that $s$ belongs to the interval $(s_{k_0-1+E},s_{k_0+2E}].$  Because the number $E$ is sufficiently large such that $\displaystyle E\ge \frac{1}{\alpha \underline{\kappa}} \ln\Big( \frac{\gamma R_1(1+NL_f \bar{\delta})^p}{\epsilon} \Big)$ and $s-s_{k_0-1}> E\underline{\kappa},$ we have 
$$
R_1 (1+NL_f \bar{\delta})^p e^{-\alpha (s-s_{k_0-1})} < \frac{\epsilon}{\gamma}.
$$
 Hence,
\begin{eqnarray*}
&& \Big\| \phi_{\zeta}(s) - \phi_{\tilde{\zeta}}(s)  \Big\|  < \displaystyle \frac{\epsilon}{\gamma} + \displaystyle \frac{N\epsilon}{\gamma} \left(  \frac{1}{\lambda} + \frac{\bar{\delta} p }{1-e^{-\lambda \psi(\omega)}}  \right) \left( 1+  \frac{NL_f(1+NL_f\bar{\delta})^p}{\alpha} + \frac{NL_f\bar{\delta}p(1+NL_f\bar{\delta})^p}{1-e^{-\alpha \psi(\omega)}} \right) \\
&& \le \epsilon.
\end{eqnarray*}
The last inequality yields $ \left\| \varphi_{\zeta}(t) - \varphi_{\tilde{\zeta}}(t)  \right\|  < \epsilon$ for $t\in [\theta_{2(k_0+E)-1}, \theta_{2(k_0+2E)}]\cap \mathbb T_0.$ Consequently, the couple  $\varphi_{\zeta}(t), \varphi_{\tilde{\zeta}}(t)$
is proximal. $\square$

The frequent separation feature of the bounded solutions of (\ref{main_eqn1}) is presented in the next lemma.

\begin{lemma} \label{timescale_freq}
Under the conditions $(C1)-(C5),$ if a pair of sequences $\zeta, \tilde{\zeta} \in \Theta$ is frequently separated, then the pair of solutions $\varphi_{\zeta}(t), \varphi_{\tilde{\zeta}}(t)\in\mathscr{A}$ is frequently $(\epsilon_0, \Delta)$-separated for some positive numbers $\epsilon_0$ and $\Delta.$
\end{lemma}

\noindent \textbf{Proof.} 
Because the pair of sequences $\zeta, \tilde{\zeta}$ is frequently separated, there exists a positive number $\bar{\epsilon}_0$ and a sequence $\left\{k_q\right\}$ of integers satisfying $k_q \to \infty$ as $q \to \infty$ such that $\left\| \zeta_{k_q} - \tilde{\zeta}_{k_q} \right\| > \bar{\epsilon}_0$ for each $q\in \mathbb N.$

Let us fix a natural number $q.$ For $s\in (s_{k_q-1}, s_{k_q}],$ the solutions $\phi_{\zeta}(s)=\varphi_{\zeta}(\psi^{-1}(s))$ and $\phi_{\tilde{\zeta}}(s)=\varphi_{\tilde{\zeta}}(\psi^{-1}(s))$ of (\ref{main_eqn3}) satisfy the relations
\begin{eqnarray*}
\phi_{\zeta} (s) = \phi_{\zeta} (s_{k_q-1}+) + \displaystyle \int^s_{s_{k_q-1}} [A \phi_{\zeta} (r) + f(\psi^{-1}(r),\phi_{\zeta}(r)) + \zeta_{k_q}] dr  
\end{eqnarray*} 
and
\begin{eqnarray*}
\phi_{\tilde{\zeta}} (s) = \phi_{\tilde{\zeta}} (s_{k_q-1}+) + \displaystyle \int^s_{s_{k_q-1}} [A \phi_{\tilde{\zeta}} (r) + f(\psi^{-1}(r),\phi_{\tilde{\zeta}}(r)) +  \tilde{\zeta}_{k_q} ] dr,  
\end{eqnarray*} 
respectively. Therefore, one can obtain that
\begin{eqnarray*}
&& \left\|\phi_{\zeta} (s_{k_q})- \phi_{\tilde{\zeta}} (s_{k_q})\right\| > \bar{\epsilon}_0 \underline{\kappa} - \left\|\phi_{\zeta} (s_{k_q-1}+)- \phi_{\tilde{\zeta}} (s_{k_q-1}+)\right\| - \displaystyle \int^{s_{k_q}}_{s_{k_q-1}} \left(\left\|A\right\|+L_f\right) \left\|\phi_{\zeta} (r)- \phi_{\tilde{\zeta}} (r)\right\| dr \\
&& \ge \bar{\epsilon}_0 \underline{\kappa} - [1+\left(\left\|A\right\|+L_f\right)\overline{\kappa}] \sup_{s\in (s_{k_q-1},s_{k_q}]}  \left\|\phi_{\zeta} (s)- \phi_{\tilde{\zeta}} (s)\right\|.
\end{eqnarray*} 
The last inequality implies that 
\begin{eqnarray*}
\sup_{s\in (s_{k_q-1},s_{k_q}]}  \left\|\phi_{\zeta} (s)- \phi_{\tilde{\zeta}} (s)\right\| > \displaystyle \frac{ \bar{\epsilon}_0 \underline{\kappa}}{2+\left(\left\|A\right\|+L_f\right) \overline{\kappa}}.
\end{eqnarray*} 
Define the number
$$
\displaystyle \Delta=\min \left\{  \frac{\underline{\kappa}}{2}, \frac{\bar{\epsilon}_0 \underline{\kappa}}{4[2+(\left\|A\right\|+L_f)\overline{\kappa}](K_0\left\|A\right\|+M_f+M_F)}  \right\}.
$$
At first, suppose that $\displaystyle\sup_{s\in (s_{k_q-1},s_{k_q}]}  \left\|\phi_{\zeta} (s)- \phi_{\tilde{\zeta}} (s)\right\| = \left\|\phi_{\zeta} (\eta)- \phi_{\tilde{\zeta}} (\eta)\right\|$ for some $\eta \in (s_{k_q-1},s_{k_q}],$ and let
\begin{eqnarray*} 
\nu_q=\left\{\begin{array}{ll}   \eta,    & \textrm{if}~  \eta \le (s_{k_q-1}+s_{k_q})/2, \\
                                \eta-\Delta,    &\textrm{if}~    \eta>(s_{k_q-1}+s_{k_q})/2, 
\end{array} \right..
\end{eqnarray*}
It can be verified for $s\in \widetilde{J}_q=[\nu_q,\nu_q+\Delta]$ that  
\begin{eqnarray*}
&& \left\|\phi_{\zeta} (s) - \phi_{\tilde{\zeta}} (s)\right\| \ge  \left\|\phi_{\zeta} (\eta) - \phi_{\tilde{\zeta}} (\eta)\right\| - \left| \displaystyle \int^s_{\eta} \left\|A\right\|  \left\|\phi_{\zeta} (r) - \phi_{\tilde{\zeta}} (r) \right\| dr \right| \\
&& - \displaystyle \left|\int^s_{\eta}  \left\|  f(\psi^{-1}(r),\phi_{\zeta}(r))  - f(\psi^{-1}(r),\phi_{\tilde{\zeta}}(r)) \right\| dr \right| - \displaystyle \left|\int^s_{\eta}  \left\|  \eta_{k_q}-\tilde{\eta}_{k_q} \right\|  dr\right| \\
&& > \displaystyle \frac{\bar{\epsilon}_0 \underline{\kappa}}{2[2+ (\left\|A\right\|+L_f)\overline{\kappa}] }. 
\end{eqnarray*} 
On the other hand, the inequality $\left\|\phi_{\zeta} (s) - \phi_{\tilde{\zeta}} (s)\right\| > \displaystyle \frac{\bar{\epsilon}_0 \underline{\kappa}}{2[2+ (\left\|A\right\|+L_f)\overline{\kappa}] }$ is true also for $s\in \widetilde{J}_q=(s_{k_q-1},s_{k_q-1}+\Delta]$ in the case that $\displaystyle\sup_{s\in (s_{k_q-1},s_{k_q}]}  \left\|\phi_{\zeta} (s)- \phi_{\tilde{\zeta}} (s)\right\| = \left\|\phi_{\zeta} (s_{k_q-1}+)- \phi_{\tilde{\zeta}} (s_{k_q-1}+)\right\|.$

Thus, $\left\|\varphi_{\zeta} (t) - \varphi_{\tilde{\zeta}} (t)\right\| > \epsilon_0$ for each $t$ from the intervals $J_q,$ $q\in \mathbb N,$ where $\epsilon_0=\displaystyle \frac{\bar{\epsilon}_0 \underline{\kappa}}{2[2+ (\left\|A\right\|+L_f)\overline{\kappa}] }$ and $J_q=\psi^{-1}  (\widetilde{J}_q ).$ Consequently, the pair $\varphi_{\zeta}(t), \varphi_{\tilde{\zeta}}(t)\in\mathscr{A}$ is frequently $(\epsilon_0, \Delta)$-separated. $\square$

The main result of the present study is mentioned in the following theorem. 

\begin{theorem}\label{main_timescale}
Assume that the conditions $(C1)-(C7)$ are fulfilled. If the map (\ref{timescale_mapF}) is Li-Yorke chaotic on $\Lambda,$ then system (\ref{main_eqn1}) is chaotic in the sense of Definition \ref{timescale_chaos_defn}.
\end{theorem}

\noindent  \textbf{Proof.}
Suppose that $\zeta=\left\{\zeta_k\right\}$ is a $p_0-$periodic solution of (\ref{timescale_mapF}) for some $p_0\in\mathbb N.$ In this case, the function $g(t,\zeta),$ which is used in the right hand side of equation (\ref{main_eqn1}), is $m\omega-$periodic, where $m=lcm \left\{p_0,p\right\}/p.$ Making use of the conditions $(C5)$ and $(C7),$ one can verify that the bounded solution $\varphi_{\zeta}(t)$ of (\ref{main_eqn1}) is $m\omega-$periodic. Therefore, (\ref{main_eqn1}) possesses an $m\omega-$periodic solution for each $m\in\mathbb N.$

Let us denote by $\Sigma$ the set consisting of bounded solutions $\varphi_{\zeta}(t)$ of (\ref{main_eqn1}) for which the initial value $\zeta_0$ of the sequence $\zeta=\left\{\zeta_k\right\}$ belongs to the scrambled set ${\mathcal S}$ of the map (\ref{timescale_mapF}). Because the set ${\mathcal S}$ is uncountable, $\Sigma$ is also uncountable. Moreover, $\Sigma$ does not contain any periodic solutions, since no periodic points of $F$ take place inside ${\mathcal S}.$  

According to the Lemmas \ref{timescale_prox} and \ref{timescale_freq}, any pair of different solutions inside $\Sigma$ is a Li-Yorke pair, i.e. $\Sigma$ is a scrambled set. Besides, Lemma \ref{timescale_freq} implies that for any solution $\varphi_{\zeta}(t) \in \Sigma$ and any periodic solution $\varphi_{\hat{\zeta}}(t) \in \mathscr{A},$ the pair $\varphi_{\zeta}(t),$ $\varphi_{\hat{\zeta}}(t)$ is frequently $(\epsilon_{0},\Delta)$-separated for some positive numbers $\epsilon_{0}$ and $\Delta.$ Consequently, system (\ref{main_eqn1}) is Li-Yorke chaotic. $\square$

In the next section, a Duffing equation on a time scale will be utilized to illustrate the theoretical results.
 
\section{An example} \label{example_time_scales}

Let us take into account the following forced Duffing equation,
\begin{eqnarray} \label{timescale_Duffing_example}
y^{\Delta \Delta} (t) + 5 y^{\Delta} (t) + \frac{35}{2} y(t) + 0.02 y^3(t) = 0.1 \cos \left(\frac{\pi}{3} t \right) + g(t,\zeta), ~ t\in \mathbb T_{0},
\end{eqnarray}  
where $\mathbb T_0 = \bigcup_{k=-\infty}^{\infty} [\theta_{2k-1}, \theta_{2k}]$ and $\theta_k= 3k+ \displaystyle \frac{1}{2} \big(1+(-1)^k \big),$ $k\in \mathbb Z.$ The function $g(t,\zeta)$ is defined through the equation $g(t,\zeta)=\zeta_k$ for $t\in [\theta_{2k-1}, \theta_{2k}],$ $k\in\mathbb Z,$ in which the sequence $\zeta=\left\{\zeta_k\right\},$ $\zeta_0 \in [0,1],$ is generated by the logistic map 
\begin{eqnarray}\label{timescale_example_map}
\zeta_{k+1}=3.9 \zeta_k (1-\zeta_k).
\end{eqnarray}
The time scale $\mathbb T_0$ satisfies the $\omega$-property with $\omega=6,$ and one can confirm that $\psi(\omega)=4$ and $\delta_k=2$ for all $k\in \mathbb Z,$ where $\delta_k=\theta_{2k+1}-\theta_{2k}.$
According to the results of the paper \cite{Li75}, the map (\ref{timescale_example_map}) possesses Li-Yorke chaos. It is worth noting that the unit interval $[0,1]$ is invariant under the iterations of the map \cite{Hale91}.
 
By using the variables $y_1=y$ and $y_2=y^{\Delta},$ equation (\ref{timescale_Duffing_example}) can be reduced to the system 
\begin{eqnarray}\label{timescale_example_system}
\begin{array}{l}
y_1^{\Delta}(t) = y_2(t), \\
y_2^{\Delta} (t) = -\displaystyle \frac{35}{2} y_1(t) - 5y_2(t) - 0.02 y_1^3(t) + 0.1 \cos \left(\frac{\pi}{3} t \right) + g(t,\zeta),
\end{array}
\end{eqnarray}
which is in the form of (\ref{main_eqn1}), where 
$$A=\left(
\begin {array}{ccc}
0 & 1 \\
\noalign{\medskip}
-\displaystyle \frac{35}{2} & -5
\end {array}
\right)$$ and
$$f(t,y_1,y_2)=\left(
\begin {array}{ccc}
0 \\
\noalign{\medskip}
-0.02y_1^3+ 0.1 \displaystyle \cos \left(\frac{\pi}{3} t \right)
\end {array}
\right).$$
One can show that 
$$
\displaystyle  e^{At}=e^{- \frac{5}{2}t}Q\left(
\begin {array}{ccc}
\cos\left(\frac{3\sqrt{5}}{2}t\right) &  - \sin\left(\frac{3\sqrt{5}}{2}t\right) \\
\noalign{\medskip}
\sin\left(\frac{3\sqrt{5}}{2}t\right) &   \cos\left(\frac{3\sqrt{5}}{2}t\right)
\end {array}
\right)Q^{-1},
$$
where
$$Q=\left(
\begin {array}{ccc}
0 & 1 \\
\noalign{\medskip}
\displaystyle \frac{3\sqrt{5}}{2} & - \displaystyle\frac{5}{2}
\end {array}
\right),$$
and the eigenvalues of the matrix $e^{4A}(I+2A)$ are inside the unit circle, where $I$ is the $2\times 2$ identity matrix. 

Due to the fact that the coefficient of the nonlinear term $y_1^3(t)$ in (\ref{timescale_example_system}) is sufficiently small, it can be numerically verified for $\zeta_0 \in [0,1]$ that the bounded solutions of system (\ref{timescale_example_system}) lie inside the region $\mathscr{D}=\left\{(y_1,y_2) \in \mathbb R^2: -0.01 \le y_1 \le 0.07,  -0.12 \le y_2 \le 0.07 \right\}.$ 
Therefore, it is reasonable to consider the dynamics of (\ref{timescale_example_system}) inside $\mathscr{D}.$

The conditions $(C5)$ and $(C6)$ hold for (\ref{timescale_example_system}) with $N=193,$ $\lambda=1.6,$ $p=1,$ $\bar{\delta}=2$ and $L_f=0.000294.$ In accordance with Theorem \ref{main_timescale}, system (\ref{timescale_example_system}) is Li-Yorke chaotic. It is worth noting that the chaoticity of the logistic map (\ref{timescale_example_map}) gives rise to the presence of chaos in (\ref{timescale_example_system}). Moreover, Lemma \ref{attractiveness_timescalepaper} implies that for a fixed solution $\zeta=\left\{\zeta_k\right\}$ of (\ref{timescale_example_map}) the unique bounded solution of (\ref{timescale_example_system}) attracts all other solutions of the system.

\begin{figure}[ht] 
\centering
\includegraphics[width=13.5cm]{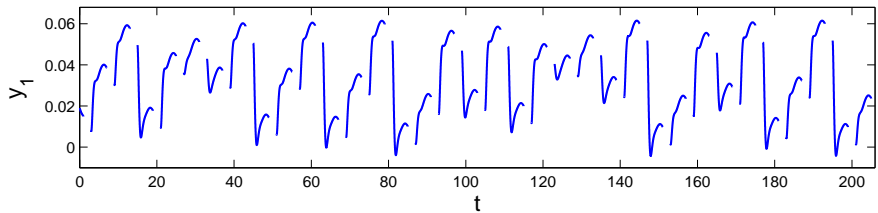}
\caption{The chaotic behavior in the solution of system (\ref{timescale_example_system}).}
\label{timescale_fig1}
\end{figure}

\begin{figure}[ht] 
\centering
\includegraphics[width=8.5cm]{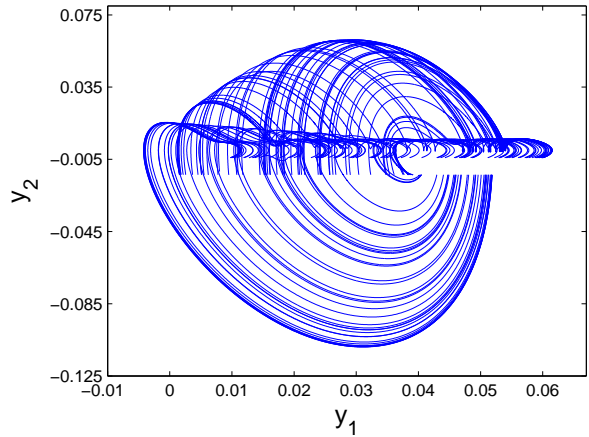}
\caption{The chaotic trajectory of system (\ref{timescale_example_system}).}
\label{timescale_fig2}
\end{figure}

Let us use the solution $\zeta=\left\{\zeta_k\right\}$ of (\ref{timescale_example_map}) with $\zeta_0=0.19$ in system (\ref{timescale_example_system}). We depict in Figure \ref{timescale_fig1} the $y_1$-coordinate of the solution of (\ref{timescale_example_system}) corresponding to the initial data $y_1(0)=0.019$ and $y_2(0)=-0.004.$ Figure \ref{timescale_fig1} supports the result of Theorem \ref{main_timescale} such that system (\ref{timescale_example_system}) possesses chaos.  Moreover, the trajectory of the same solution in the $y_1-y_2$ plane is represented in Figure \ref{timescale_fig2}, which reveals the existence of a chaotic attractor in the dynamics of (\ref{timescale_example_system}).

\section{Conclusion} \label{time_scales_conclusion}
 
We rigorously prove the existence of chaos in dynamic equations on time scales, where the right hand side of the equations depends on a chaotic map. The reduction technique to impulsive differential equations presented in the paper \cite{Akhmet06} is used in our investigations. A mathematical description of chaos in the sense of Li-Yorke is provided for DETS, and the ingredients of the Li-Yorke chaos, proximality and frequent separation, are theoretically proved. The results can be used to obtain chaotic mechanical systems and electrical circuits on time scales without any restriction in the dimension.

\section*{Acknowledgments}

The authors wish to express their sincere gratitude to the referees for the helpful criticism and valuable suggestions, which helped to improve the paper significantly.

The second author is supported by the 2219 scholarship programme of T\"{U}B\.{I}TAK, the Scientific and Technological Research Council of Turkey.


\begin{thebibliography}{30}

\bibitem{Akhmet06} Akhmet, M. U. \& Turan, M. [2006] ``The differential equations on time scales through impulsive differential equations,'' {\it Nonlinear Analysis} \textbf{65}, 2043--2060.

\bibitem{Akhmet09} Akhmet, M. U. \& Turan, M. [2009] ``Differential equations on variable time scales,'' {\it Nonlinear Analysis} \textbf{70}, 1175--1192.

\bibitem{Akh5} Akhmet, M. U. [2009a] ``Devaney's chaos of a relay system,'' {\it Commun. Nonlinear Sci. Numer. Simulat.} \textbf{14}, 1486--1493.

\bibitem{Akh2} Akhmet, M. U. [2009b] ``Li-Yorke chaos in the impact system,'' {\it J. Math. Anal. Appl.} \textbf{351}, 804--810.

\bibitem{Akh4} Akhmet, M. U. [2009c] ``Dynamical synthesis of quasi-minimal sets,'' {\it Int. J. Bifurcation and Chaos} \textbf{19}, 2423--2427. 

\bibitem{Akh1} Akhmet, M. [2010] {\it Principles of Discontinuous Dynamical Systems}, (Springer, New York). 

\bibitem{Akh7} Akhmet, M. U. \& Fen, M. O. [2012a] ``Chaotic period-Doubling and OGY control for the forced Duffing equation,'' {\it Commun. Nonlinear Sci. Numer. Simulat.} \textbf{17}, 1929--1946. 

\bibitem{Akh9} Akhmet, M. U. \& Fen, M. O. [2012b] ``Chaos generation in hyperbolic systems,'' {\it Interdiscip. J. Discontinuity, Nonlinearity, and Complexity} \textbf{1}, 353--365.
 
\bibitem{Akh8} Akhmet, M. U. \& Fen, M. O. [2013] ``Replication of chaos,'' {\it Commun. Nonlinear Sci. Numer. Simulat.} \textbf{18}, 2626--2666. 
 
\bibitem{Akh14} Akhmet, M. U. \& Fen, M. O. [2014] ``Entrainment by chaos,'' {\it J. Nonlinear Sci.} \textbf{24}, 411--439.

\bibitem{Akin03} Akin, E. \& Kolyada, S. [2003] ``Li-Yorke sensitivity,'' {\it Nonlinearity} \textbf{16}, 1421--1433.

\bibitem{Andersson94} Andersson, K. G. [1994] ``Poincar\'{e}'s discovery of homoclinic points,'' {\it Archive for History of Exact Sciences} \textbf{48}, 133--147.

\bibitem{Aulbach01} Aulbach, B. \& Kieninger, B. [2001] ``On three definitions of chaos,'' {\it Nonlinear Dynamics and Systems Theory} \textbf{1}, 23--37.

\bibitem{Barrio14} Barrio, R., Martinez, M. A., Serrano, S. \& Shilnikov, A. [2014] ``Macro- and micro-chaotic structures in the Hindmarsh-Rose model of bursting neurons,'' {\it Chaos} \textbf{24}, 023128.

\bibitem{Blanchard02} Blanchard, F., Glasner, E., Kolyada, S. \&  Maass, A. [2002] ``On Li-Yorke pairs,'' {\it J. Reine Angew. Math.} \textbf{2002}, 51--68.

\bibitem{Bohner01} Bohner, M. \& Peterson, A. [2001] {\it Dynamic Equations on Time Scales: An Introduction with Applications}, (Birkh\"{a}user, Boston).

\bibitem{Brown93} Brown, R. \& Chua, L. [1993] ``Dynamical synthesis of Poincar\'{e} maps,'' {\it Int. J. Bifurcation and Chaos} {\bf 3}, 1235--1267.

\bibitem{Brown96} Brown, R. \& Chua, L. [1996] ``From almost periodic to chaotic: the fundamental map,'' {\it Int. J. Bifurcation and Chaos} {\bf 6}, 1111--1125.

\bibitem{Brown97} Brown, R. \& Chua, L. [1997] ``Chaos: generating complexity from simplicity,'' {\it Int. J. Bifurcation and Chaos} {\bf 7}, 2427--2436.

\bibitem{Brown01} Brown, R., Berezdivin, R. \& Chua, L. [2001] ``Chaos and complexity,'' {\it Int. J. Bifurcation and Chaos} {\bf 11}, 19--26.

\bibitem{Cartwright1} Cartwright, M. \& Littlewood, J. [1945] ``On nonlinear differential equations of the second order I: The equation $\ddot{y}- k(1 - y^2)'y + y = bk cos(\lambda t + a),$ $k$ large,'' {\it J. London Math. Soc.} \textbf{20}, 180--189.

\bibitem{Feckan11} Fe\v ckan, M. [2011] {\it Bifurcation and Chaos in Discontinuous and Continuous Systems}, (Springer-Verlag, Heidelberg).

\bibitem{Grebogi97} Grebogi, C. \& Yorke, J. A. [1997] {\it The Impact of Chaos on Science and Society}, (United Nations University Press, Tokyo).

\bibitem{Guirao05} Guirao, J. L. G. \& Lampart, M. [2005] ``Li and Yorke chaos with respect to the cardinality of the scrambled sets,'' {\it Chaos, Solitons and Fractals} \textbf{24}, 1203--1206.

\bibitem{Hale91} Hale, J. \& Ko\c{c}ak, H. [1991] {\it Dynamics and Bifurcations}, (Springer-Verlag, New York).

\bibitem{Hilger88} Hilger, S. [1988] ``Ein Ma\ss kettenkalk\"{u}l mit Anwendung auf Zentrumsmanningfaltigkeiten,'' {\it PhD thesis}, Universit\"{a}t W\"{u}rzburg.

\bibitem{Horn92} Horn, R. A. \& Johnson, C. R. [1992] {\it Matrix Analysis}, (Cambridge University Press, United States of America).

\bibitem{Kloeden06} Kloeden, P. \& Li, Z. [2006] ``Li-Yorke chaos in higher dimensions: a review,'' {\it Journal of Difference Equations and Applications} \textbf{12}, 247--269.

\bibitem{Kolyada04} Kolyada, S. F. [2004] ``Li-Yorke sensitivity and other concepts of chaos,'' {\it Ukrainian Mathematical Journal} \textbf{56}, 1242--1257.

\bibitem{Kuchta} Kuchta, M. \& Sm\'{i}tal, J. [1989] ``Two point scrambled set implies chaos,'' {\it European Conference on Iteration Theory (ECIT 87)}, (World Sci. Publishing, Singapore), pp. 427--430. 

\bibitem{Lakshmikantham96} Lakshmikantham, V., Sivasundaram, S. \& Kaymakcalan, B. [1996] {\it Dynamic Systems on Measure Chains}, (Kluwer Academic Publishers, Netherlands).

\bibitem{Lakshmikantham02} Lakshmikantham, V. \& Vatsala, A. S. [2002] ``Hybrid systems on time scales,'' {\it J. Comput. Appl. Math.} \textbf{141}, 227--235.

\bibitem{Lakshmikantham06} Lakshmikantham, V. \& Devi, J. V. [2006] ``Hybrid systems with time scales and impulses,'' {\it Nonlinear Analysis} \textbf{65}, 2147--2152.

\bibitem{Levinson} Levinson, N. [1949] ``A second order differential equation with singular solutions,'' {\it Ann. of Math.} \textbf{50}, 127--153.

\bibitem{Li75} Li, T. Y. \& Yorke, J. A. [1975] ``Period three implies chaos,'' {\it Amer. Math. Monthly} \textbf{87}, 985--992.

\bibitem{Li93} Li, S. [1993] ``$\omega$-chaos and topological entropy,'' {\it Transactions of the American Mathematical Society} \textbf{339}, 243--249.

\bibitem{PLi07} Li, P., Li, Z., Halang, W.A. \& Chen, G. [2007] ``Li-Yorke chaos in a spatiotemporal chaotic system,'' {\it Chaos, Solitons and Fractals} \textbf{33}, 335--341.

\bibitem{Lorenz63} Lorenz, E. N. [1963] ``Deterministic nonperiodic flow,'' {\it J. Atmos. Sci.} \textbf{20}, 130--141.

\bibitem{AlbertLuo} Luo, A. C. J. [2014] {\it Toward Analytical Chaos in Nonlinear Systems}, (John Wiley \& Sons, United Kingdom).

\bibitem{Marotto78} Marotto, F. R. [1978] ``Snap-back repellers imply chaos in $\mathbb R^{n}$,'' {\it J. Math. Anal. Appl.} \textbf{63}, 199--223.

\bibitem{Owens13} Owens, B. A. M., Stahl, M. T., Corron, N. J., Blakely, J. N. \& Illing, L. [2013] ``Exactly solvable chaos in an electromechanical oscillator,'' {\it Chaos} \textbf{23}, 033109.

\bibitem{Samolienko95} Samoilenko, A. M. \& Perestyuk, N. A. [1995] {\it Impulsive Differential Equations}, (World Scientific, Singapore).

\bibitem{Shi04} Shi, Y. \& Chen, G. [2004] ``Chaos of discrete dynamical systems in complete metric spaces,'' {\it Chaos, Solitons \& Fractals} \textbf{22}, 555--571. 

\bibitem{Shi05} Shi, Y. \& Chen, G. [2005] ``Discrete chaos in Banach spaces,'' {\it Science in China, Ser. A: Mathematics} \textbf{48}, 222--238. 

\bibitem{Thamilmaran04} Thamilmaran, K., Lakshmanan, M. \& Venkatesan, A. [2004] ``Hyperchaos in a modified canonical Chua's circuit,'' {\it Int. J. Bifurcation and Chaos} \textbf{14}, 221--243.

\bibitem{Tisdell08} Tisdell, C. C. \& Zaidi, A. [2008] ``Basic qualitative and quantitative results for solutions to nonlinear, dynamic equations on time scales with an application to economic modelling,'' {\it Nonlinear Analysis} \textbf{68}, 3504--3524.

\bibitem{Ueda78} Ueda, Y. [1978] ``Random phenomena resulting from non-linearity in the system described by Duffing's equation,'' {\it Trans. Inst. Electr. Eng. Jpn.} \textbf{98A}, 167--173.

\bibitem{Zhang10} Zhang, J., Fan, M. \& Zhu, H. [2010] ``Periodic solution of single population models on time scales,'' {\it Mathematical and Computer Modelling} \textbf{52}, 515--521.




\end{thebibliography}
\end{document}